\documentstyle[prd,aps,epsf]{revtex}
\begin{document}
\title{Fermi Coordinates of an Observer Moving in a Circle in Minkowski Space:\\
Apparent Behavior of Clocks}
\author{Thomas B. Bahder}
\address{Army Research Laboratory \\
2800 Powder Mill Road \\
Adelphi, Maryland USA  20783}

\date{\today}
\maketitle


\begin{abstract}
Coordinate transformations are derived from global Minkowski
coordinates to the Fermi coordinates of an observer moving in a
circle in Minkowski space-time. The metric for the Fermi
coordinates is calculated directly from the tensor transformation
rule. The behavior of ideal clocks is examined from the observer's
reference frame using the Fermi coordinates.  A complicated
relation exists between Fermi coordinate time and proper time on
stationary clocks (in the Fermi frame) and between proper time on
satellite clocks that orbit the observer. An orbital Sagnac-like
effect exists for portable clocks that orbit the Fermi coordinate
origin. The coordinate speed of light is isotropic but varies with
Fermi coordinate position and time. The magnitudes of these
kinematic effects are computed for parameters that are relevant to
the Global Positioning System (GPS) and are found to be
small; however, for future high-accuracy time transfer systems, these
effects may be of significant magnitude.
\end{abstract}


\section{Introduction}

During the past several decades there have been significant
advances in the accuracy of satellite-based navigation and time
transfer systems~\cite{ParknisonBeardReview}. The most recent
example of such a system is the Global Positioning System (GPS),
which is run by the U. S. Department of
Defense~\cite{ParkinsonGPSReview,Kaplan96,Hofmann-Wellenhof93}.
The GPS is based on a network of 24 atomic clocks in space, from
which a user can receive signals and compute his position and
time. The order of magnitude of the accuracy of the GPS currently
is ten meters and twenty nanoseconds~\cite{NRCReport}. Future
navigation and time transfer systems are being contemplated that
will have several orders of magnitude better position accuracy  and
sub-nanosecond time accuracy~\cite{AsbyAllan96,PetitWolf94,WolfPetit95}.

Over satellite to ground distances, precise measurements must be
interpreted within the framework of a relativistic
theory~\cite{Guinot97}. The design of high-accuracy navigation and
time transfer systems depends on detailed modeling of the
measurement process. The comparison of experiment with theory in
general relativity is more subtle than in special relativity or
other areas of physics for two reasons. First, due to the
possibility of making coordinate gauge transformations, often it
is not clear what is a measurable quantity and what is an artifact
of the coordinate system~\cite{SchiffXX,Bergmann88}. Second, all
measurements must in-principle be interpreted in the comoving
frame of the experimental apparatus~\cite{Synge1960}. In the case
of the GPS, even theoretical relativistic treatments do not
interpret the measured observable in the comoving frame of the
apparatus, see for example Ref.\ ~\cite{Schwarze93}.

The principle reason for investigating in detail relativistic
effects is to improve the current accuracy of GPS and to create
future time transfer and navigation systems that have several
orders of magnitude better accuracy.  At the present time, it is
well-known that small anomalies exist in position and time
computed from GPS data. The origin of these anomalies is not
understood. In particular, GPS time transfer data from the
\mbox{U. S.} Naval Observatory indicates that GPS time is periodic
with respect to the Master Clock, which is the most accurate
source of official time for the \mbox{U. S.} Department of
Defense~\cite{Weiss87,Bahder97,Bahder98}. Furthermore, other
anomalies have been found in Air Force monitor station data that
are not understood at present~\cite{AlleyVanFlandernXX}.  The work
below is a first step in revisiting from a first principles
approach some of the physics that has gone into the design of the
GPS .

A theoretical framework, known as the tetrad formalism, has been
developed for relating measurements to relativistic
theory\cite{Synge1960,Fermi22,Walker,Pirani57}. However, due to a
historical lack of accurate measurements over satellite to ground
distances, in most cases it has not been necessary to apply the
tetrad formalism in the area of satellite-based systems.  This
situation is rapidly changing due to improved measurement
technology\cite{Guinot97,Fukushima,AshbyAcceleration,AshbyCrosLinkRanging}.

In the tetrad formalism, measurements are interpreted as
projections of tensors on the tetrad basis vectors. Consequently,
these projections are invariant quantities (scalars) under
transformation of the space-time coordinates. However, these
projections do depend on the world line of the observer and the
choice of local Cartesian axes used by the
observer\cite{Pirani57,Synge1960}. The need for the tetrad
formalism to relate experiment to theory, and the problem of
measurable quantities in general relativity, are well discussed by
Synge~\cite{Synge1960}, Soffel~\cite{Soffel89} and
Brumberg~\cite{Brumberg91}.

The tetrad formalism was initially investigated for the case of
inertial observers, which move on
geodesics\cite{Fermi22,Walker,Pirani57,ManasseMisner,Li79,AsbyBertotti}.
However, many users of geodetic satellite systems, are
terrestrially based observers, or are based in non-inertial
platforms, such as missiles.  The theory for the case of
non-inertial observers has been investigated by
Synge\cite{Synge1960}, who considered the case of non-rotating
observers moving along a time-like world line, and by
others\cite{Li78,Ni78,Li79a,Nelson87,Nelson90}, who considered
accelerated, rotating observers.  For arbitrary observer motions,
the effects are indeed very complicated, and the general theory
gives limited physical insight. For this reason, in this work, I
work out in detail a simple model problem. I treat the particular
case of an observer moving in a circle in Minkowski space. I use
the tetrad formalism\cite{Synge1960} to obtain the transformation
from an inertial frame in Minkowski space, to the accelerated
frame of reference of an observer moving in a circle. The observer
Fermi-Walker transports his tetrad basis vectors. The case
presented here corresponds to an observer kept in a circular
trajectory by rockets, \mbox{ i.\ e.\ ,} he is not on a geodesic.
The explicit results presented below form a particular case of the
general results obtained by Synge\cite{Synge1960}, specialized to
zero gravitational field and circular motion.  The situation
treated here also corresponds to a reference frame that has a
constant orientation with respect to the distant
stars\cite{SyngeNote1}, such as an Earth-centered  inertial
reference frame used in GPS satellite computations. The results
are given here as expansions in small velocity of the observer,
compared with speed of light.  However, the coordinate
transformation equations are not expanded in a power series in
time, and  hence, they are valid for arbitrary long times.

This work explores the apparent behavior of ideal clocks as
observed from a Fermi coordinate frame of reference of an observer
moving in a circle in flat space-time. The Fermi coordinate frame
is the closest type of non-rotating frame to that which is used in
experiments and in GPS. In particular, a Fermi coordinate frame is
a non-rotating frame of reference, where Coriolis forces are
absent, and light travels essentially along a straight
line\cite{Synge1960}. The observer's Fermi frame in this work is
analogous to the Earth-centered inertial (ECI) frame used in GPS.
The ECI frame moves with the Earth but maintains its orientation
with respect to the distant stars.

This work neglects all gravitational field effects that are
significant for the actual behavior of GPS clocks. As such, this
work treats only the kinematic effects associated with the
circular motion of the Fermi coordinate frame about a central
point (analogous to the sun's position). Since gravitational
fields are neglected in this work, the observer's tetrad is
Fermi-Walker transported along a time-like world line (a circle in
3-d space). In the case of GPS, the origin of the ECI frame
follows a geodesic path (the Earth's orbit). Therefore, the
observer's Fermi frame in this work is an accelerated,
non-rotating frame, while the ECI frame is in free fall. As
remarked above, the effects on clocks are in general very
complicated, even without the gravitational field. For this
reason, I work out in detail the specific case of an observer in
circular motion in Minkowski space-time, and I apply the results
to parameters that are relevant to the GPS.

\section{Fermi-Walker Transport Differential Equations}

\begin{figure}[htbp]
\centerline{\epsfsize=6.0cm \epsfbox{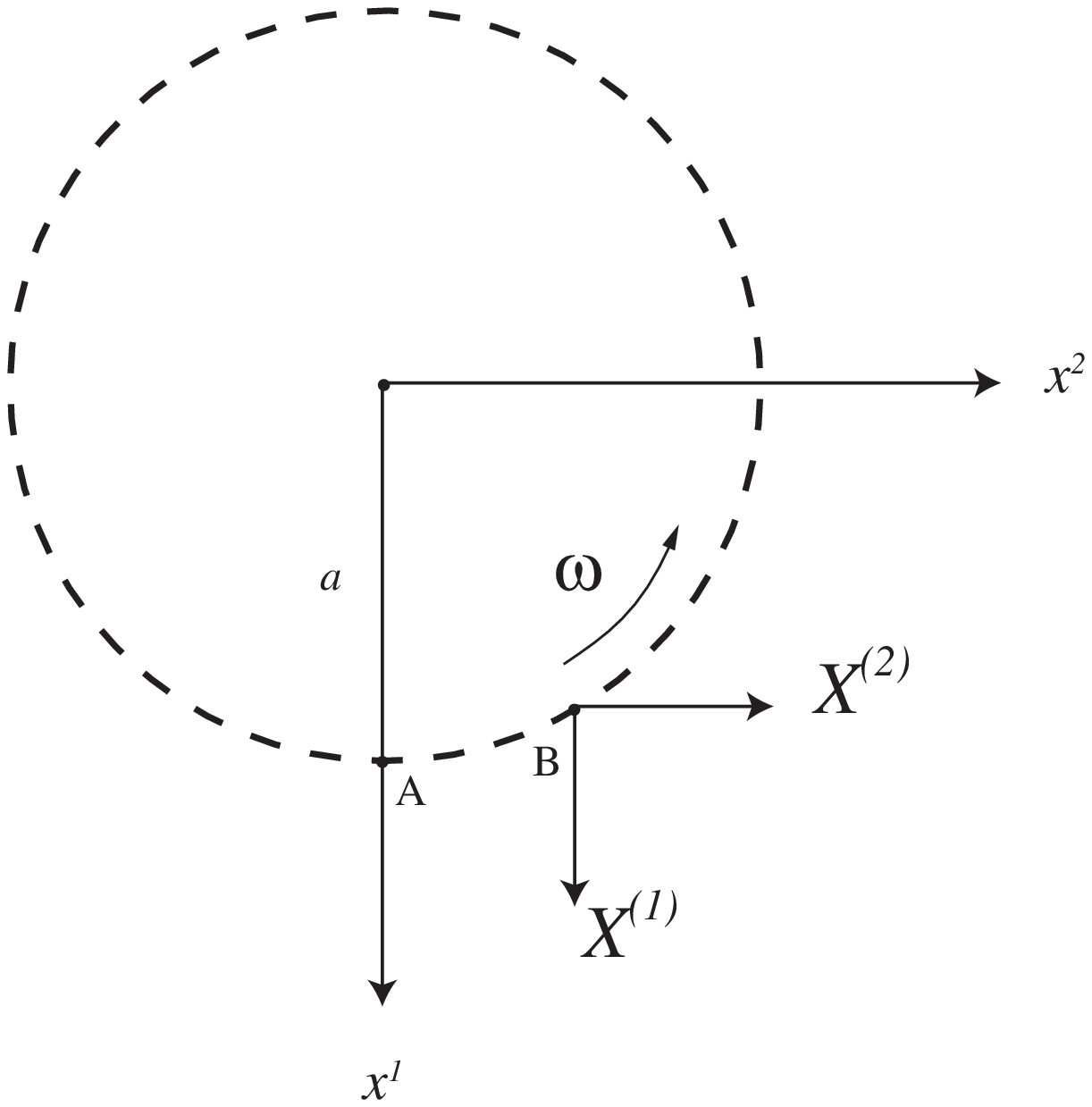}}
\caption{The observer's circular 3-d path is shown against the
background of the global inertial Minkowski coordinates $x^i$. The
position of the observer (origin of Fermi coordinates) and the
orientation of the Fermi coordinate axes $X^{(a)}$ are shown
schematically at a given time.} \protect
\label{ObserverMotionFigure}
\end{figure}

Consider a time-like world line $C$ of an observer moving in a circle
in 3-dimensional (3-d) space given by
\begin{eqnarray}
x^0 & = & x_0^0 + u  \label{worldLine0} \\
x^1 & = & x_0^1 + a \cos \left( \omega u /c \right) \label{worldLine1}\\
x^2 & = & x_0^2 + a \sin \left( \omega u /c \right) \label{worldLine2} \\
x^3 & = & x_0^3 \label{worldLine3}
\end{eqnarray}
where $x_0^0$, $x_0^1$, $x_0^2$, and $x_0^3$ are constants, $u$ is
a parameter along the world line,  and $c$ is the speed of light
in an inertial frame in vacuum. The observer travels in a circle
in the $x-y$ plane. See Figure \ref{ObserverMotionFigure}. Along
his world line, the observer travels with a four velocity
\begin{equation}
u^i =\frac{d x^i}{ds} \label{fourVelocity}
\end{equation}
and has an acceleration
\begin{equation}
w^i =\frac{\delta u^i}{\delta s} =
\frac{d u^i}{d s} + \Gamma^i_{j k} u^j u^k  \label{fourAcceleration}
\end{equation}
where $\Gamma^i_{j k}$ is the affine connection.
From the normalization of the 4-velocity, $u^i u_i = -1$, I obtain
the relation between the arc length, $s=c \tau$, where $\tau$ is
the proper time, and the parameter $u$,
\begin{equation}
  u = \gamma s
\end{equation}
where $\gamma = (1- \nu^2 )^{-1/2}$. The parameter $\nu$ is the
dimensionless velocity, given by
\begin{equation}
\nu = \frac{a \omega}{c} \label{nuDef}
\end{equation}
I use the convention that Roman indices take the values
$i=0,1,2,3$ and Greek indices take values $\alpha=1,2,3$, see the
Appendix for the conventions used in this work.

In Minkowski space, the affine connection components all vanish,
$\Gamma^i_{j k}=0$.  Hence, the explicit expressions for the
4-velocity and acceleration are:
\begin{equation}
u^i (s) = \gamma  \left(1,  -\frac{a \omega}{c} \sin ( \frac{\gamma \omega s}{c} ),
\frac{a \omega}{c} \cos ( \frac{\gamma \omega s}{c}), 0
\right)
\label{fourVelocity2}
\end{equation}
\begin{equation}
w^i = -a \gamma^2 \left( \frac{\omega}{c}\right)^2
\left(0,  \cos ( \frac{\gamma \omega}{c} s), \sin (\frac{\gamma \omega}{c} s), 0
\right)
\label{fourAcceleration2}
\end{equation}

\begin{figure}[htbp]
\centerline{\epsfsize=6.0cm \epsfbox{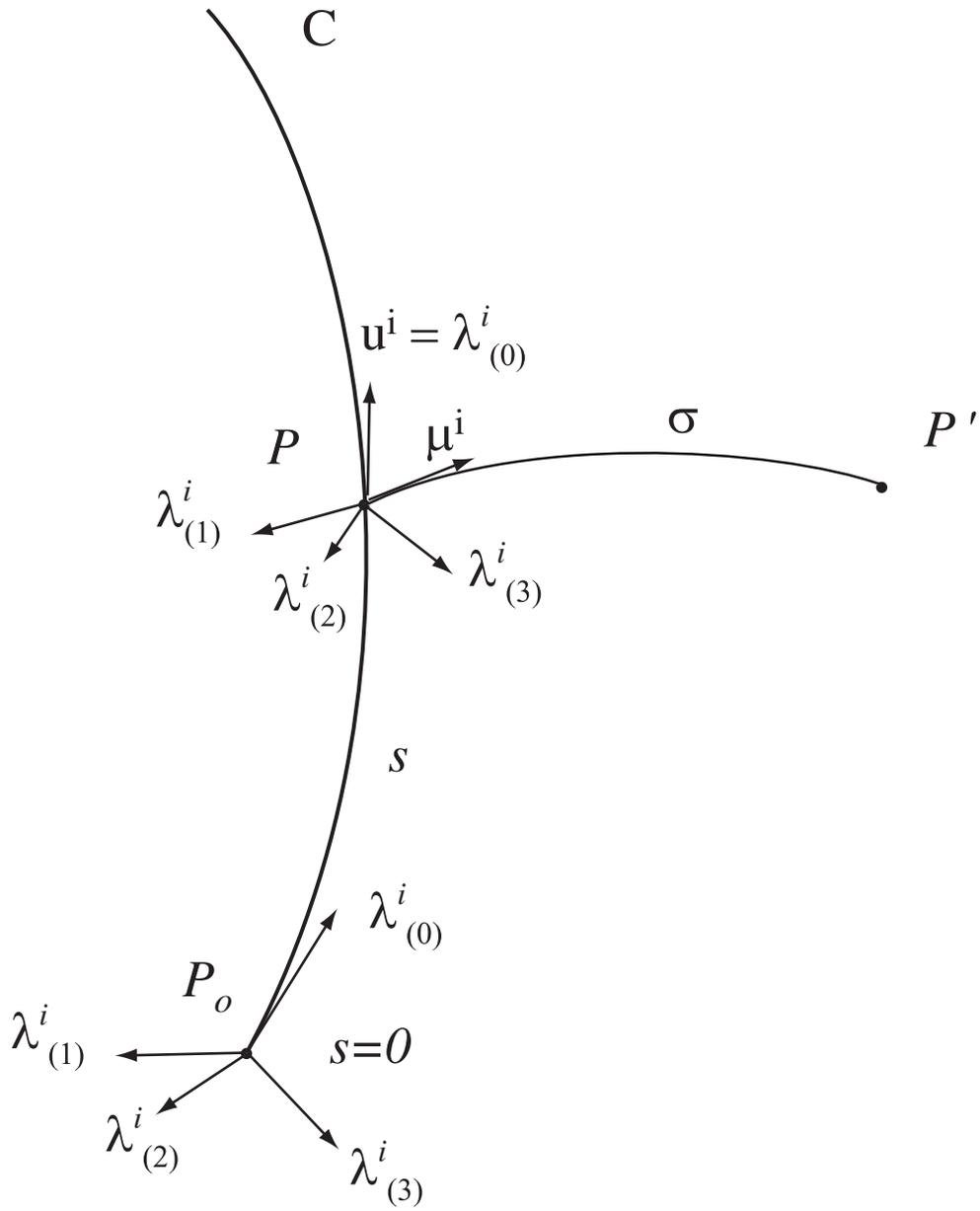}} \caption{The
observer's world line $C$ is shown with the initial tetrad basis
vectors $\lambda^i_{\alpha}$ at $s=0$ at point $P_0$. The Fermi
transported tetrad basis vectors at finite proper time $s$ are
shown at point $P$. } \protect \label{ObserverWorldLineFigure}
\end{figure}

As the observer moves on the time-like world line $C$, he carries
with him an ideal clock and three gyroscopes.  I use the notation
$x^i$ for Minkowski space coordinates. At some initial coordinate
time $x^0$, the observer is at point $P_0$ at proper time
$\tau=0$, corresponding to $s=0$. On his world line, the observer
carries with him three orthonormal basis vectors,
$\lambda_{(\alpha)}^{(i)}$, where $\alpha=1,2,3$, labels the
vectors and $i=0,1,2,3$ labels the components of these vectors in
Minkowski coordinates. These vectors form the basis for his
measurements~\cite{Pirani57}. See
Figure~\ref{ObserverWorldLineFigure}. The orientation of each
basis vector is held fixed with respect to each of the gyroscope's
axis of rotation~\cite{MTW}. The fourth basis vector is taken to
be the observer 4-velocity, $\lambda_{(0)}^{(i)} = v^i$. The four
unit vectors $\lambda_{(a)}^{(i)}$ form our
tetrad\cite{Synge1960}, an orthonormal set at $P_0$, which satisfy
\begin{equation}
g_{ij} \, \lambda^i_{(a)} \, \lambda^j_{(b)} = \delta_{a b}
\label{initialTetrad}
\end{equation}
where the $\delta_{ij}=1$ if $i=j,$ and zero otherwise.

At a later time $s=c \tau$, the observer is at a point $P$. The
observer's orthonormal set of basis vectors are related to his
tetrad basis at $P_0$, by Fermi-Walker transport. Fermi-Walker
transport preserves the lengths and relative angles of the
transported vectors. Furthermore, Fermi-Walker transport of basis
vectors along the observer's world line corresponds to keeping the
vectors as parallel as possible by keeping their orientation
constant relative to the spin axes of the gyroscopes~\cite{MTW}.
For a given vector, with contravariant components $f^i$ in
Minkowski space, its components at $P$ are related to its
components at $P_0$ by the Fermi-Walker transport differential
equations~\cite{Synge1960}
\begin{equation}
\frac{\delta f^i}{\delta s} = W^{ij} f_j
\label{FWtransport}
\end{equation}
where
\begin{equation}
W^{ij}= u^i w^i - w^i u^j
\label{FWtensor}
\end{equation}
When we use Eq.\ (\ref{FWtensor}) to transport a vector $f^i$ that
is orthogonal to the 4-velocity, $ u^i f_i = 0 $, the second term
in Eq.\ (\ref{FWtensor}) does not contribute. We refer to
transport of such space-like basis vectors as Fermi transport, and
$ W^{i j} \rightarrow \tilde{W}^{ij}= u^i w^i $.  For an arbitrary
vector $f^i$ perpendicular to the 4-velocity, the explicit form of
the Fermi transport differential equations for the world line in
Eqs.\ (\ref{worldLine0})--(\ref{worldLine3}) is
\begin{eqnarray}
\frac{d f^0}{d \xi} & = &
-\gamma^3 \nu \cos\left(\gamma \xi \right) f^1 - \gamma^3 \nu \sin
\left( \gamma \xi\right) f^2 \label{FDiffEq0} \\
\frac{d f^1}{d \xi} & = & \gamma^3 \nu^2 \cos \left(\gamma \xi
\right) \sin\left(\gamma \xi \right) f^1 + \gamma^3 \nu^2 \sin^2
\left(\gamma \xi \right) f^2  \label{FDiffEq1}\\
 \frac{d f^2}{d \xi} & = &
-\gamma^3 \nu^2 \cos^2 \left(\gamma \xi \right)  f^1 - \gamma^3
\nu^2 \cos \left(\gamma \xi \right) \sin \left(\gamma \xi \right)
f^2 \label{FDiffEq2}\\
\frac{d f^3}{d \xi} & = & 0 \label{FDiffEq3}
\end{eqnarray}
where the components $f^i$ are functions of $\xi$ and the
dimensionless proper time is given by
\begin{equation}
\xi = \frac{\omega}{c}s
\end{equation}
The differential equations in Eqs.
(\ref{FDiffEq0})-(\ref{FDiffEq3}) are nothing more than the
equations that describe the Thomas precession of an electron's
spin vector as it moves in a circular orbit around the
nucleus~\cite{Weinberg72}. The solution of Eq.\
(\ref{FDiffEq0})--(\ref{FDiffEq3}) is given by~\cite{MTW}
\begin{eqnarray}
f^0 & = & -\gamma \nu A \cos \left(\gamma^2\xi+\alpha\right)+\beta \label{fDef0} \\
f^1 & = & A \cos \left(\gamma \xi\right) \cos \left(\gamma^2 \xi
+\alpha \right) + A \gamma \sin\left(\gamma \xi \right)
\sin\left(\gamma^2 \xi + \alpha \right) \label{fDef1} \\
f^2 & = & A \sin \left(\gamma \xi\right) \cos \left(\gamma^2 \xi
+\alpha \right) - A \gamma \cos\left(\gamma \xi \right)
\sin\left(\gamma^2 \xi + \alpha \right) \label{fDef2} \\
f^3 & = & \delta  \label{fDef3}
\label{fCompSolution}
\end{eqnarray}
where $A$, $\alpha$, $\delta$ and $\beta$ are integration
constants.

\section{Determination of the Tetrad}

At the point $P_0$, the tetrad basis vectors satisfy Eq.\
(\ref{initialTetrad}). These equations constitute 12 relations
(since Eq.\ (\ref{initialTetrad}) is symmetric in $\alpha$ and
$\beta$) for 16 components $\lambda^i_{(a)}$. However, the zeroth
member of the tetrad is the 4-velocity, which is assumed known
\begin{equation}
\lambda^i_{(0)} = u^i =\gamma \left(1,v^1,v^2,v^3   \right)
\label{zerothLambda}
\end{equation}
and is given by Eq.\ (\ref{fourVelocity2}).
Equation (\ref{initialTetrad}) can be rewritten as two equations
\begin{equation}
g_{i j} \lambda^i_{(a)} \lambda^j_{(0)} =0, \; \; \; \; a=1,2,3
\label{zerothOrthog}
\end{equation}
\begin{equation}
g_{i j} \lambda^i_{(\alpha)} \lambda^j_{(\beta)} = \delta_{a b},
\; \; \; \; a=1,2,3 \label{spaceOrthog}
\end{equation}
Substituting the explicit form of the Minkowski metric (see the
Appendix) into Eq.\ (\ref{zerothOrthog}), gives a condition on the
zeroth components of a tetrad vector if the spatial components are
known (Greek indices take values $\alpha$,$\kappa=$1,2,3)
\begin{equation}
\lambda^0_{(\alpha)} = \delta_{\beta \kappa} v^\beta \lambda^\kappa_{(\alpha)}
\label{zerothLambda2}
\end{equation}
Then use of Eq.\ (\ref{zerothLambda2}) to eliminate the time
components $\lambda^0_{(\alpha)}$ from Eq.\ (\ref{spaceOrthog}),
leads to an orthogonality relation for only the spatial components
of the tetrad vectors
\begin{equation}
\delta_{\mu \nu} \lambda^\mu_{(\alpha)}  \lambda^\nu_{(\beta)}
- \delta_{\mu \nu} \delta_{\kappa \gamma} v^\mu v^\kappa \lambda^\nu_{(\alpha)}
\lambda^\gamma_{(\beta)}= \delta_{\alpha \beta}
\label{lambdaSpaceOrthog}
\end{equation}

The general idea is to solve Eq.\ (\ref{lambdaSpaceOrthog}) for
the spatial components of the tetrad at $s=0$, and then to
substitute the spatial components into Eq.\ (\ref{zerothLambda2})
to obtain the time components of each tetrad vector.  Thus, I
obtain the tetrad vectors at the initial time corresponding to
$s=0$.  For each tetrad vector, these components are then used as
initial conditions in Eq.\ (\ref{fCompSolution} )to determine the
tetrad components at point $P$ at finite $s$.

The exact solution of Eq.\ (\ref{lambdaSpaceOrthog}) can be
obtained by noting that for $v^\alpha=0$, the spatial components
of the tetrad vectors are orthonormal, and the solution is given
by
\begin{equation}
\lambda^\mu_{(\alpha)}= \delta^\mu_ \alpha
\label{initialLambdaV}
\end{equation}
I assume that the tetrad unit vectors are approximately oriented
parallel to the observer's local Cartesian $x$, $y$ and $z$ axes.
For small $v^\alpha << 1$, the solution of Eq.\
(\ref{lambdaSpaceOrthog}) can be obtained as a triple power series
in $v^\alpha$, by iteration, starting with Eq.\
(\ref{initialLambdaV}) as the first approximation. Having solved for the
tetrad components as a power series, it is easy to guess the exact
solution to be
\begin{eqnarray}
\lambda^0_{(\alpha)} & = & \gamma \delta_{\alpha \beta} v^\beta
\label{initialTetrad2a}\\
\lambda^\mu_{(\alpha)}  & = &
\delta^\mu_\alpha + \frac{\gamma-1}{\nu^2} \, \delta_{\kappa
\alpha} v^\kappa v^\mu \label{initialTetrad2b}\\
\lambda^i_{(0)} & = u^i = & \gamma
\left(1,v^1,v^2,v^3 \right) \label{initialTetrad2}
\end{eqnarray}
where the components of dimensionless velocity $v^\alpha$ are
given in Eq.\ (\ref{fourVelocity2}).

Equations (\ref{initialTetrad2a})--(\ref{initialTetrad2}) give the
components of the observer's tetrad basis vectors at point $P_0$.
At point $P$, the arc length is $s>0$, and Eqs.\
(\ref{initialTetrad2a})--(\ref{initialTetrad2}) give the initial
components for each tetrad vector. These initial components are
used in the general solution of a Fermi-Walker transported vector,
given in   Eq.\ (\ref{fCompSolution}).  The tetrad components at
point $P$ at finite $s$ are given by

\begin{eqnarray}
\lambda_{(0)}^i & = &
\left\{  \gamma, \; \; - \gamma \,\nu \,\sin (\gamma \,\xi )  , \; \;
\gamma \,\nu \,\cos (\gamma \,\xi ), \; \;  0  \right\} \label{lam0} \\
 & &  \nonumber \\
\lambda_{(1)}^i & = & \left\{  - \gamma \,\nu \,\sin ({\gamma^2}\,\xi),
  \cos (\gamma \,\xi )\,\cos ({{\gamma }^2}\,\xi ) +
   \gamma \,\sin (\gamma \,\xi )\,\sin ({{\gamma }^2}\,\xi ),
   \cos ({{\gamma }^2}\,\xi )\,\sin (\gamma \,\xi ) -
   \gamma \,\cos (\gamma \,\xi )\,\sin ({{\gamma }^2}\,\xi ),
   0 \right\}   \label{lam1} \\
 & & \nonumber \\
\lambda_{(2)}^i & = &
   \left\{  \gamma \,\nu \,\cos ({{\gamma }^2}\,\xi ),
  -\gamma \,\cos ({{\gamma }^2}\,\xi )\,\sin (\gamma \,\xi )  +
   \cos (\gamma \,\xi )\,\sin ({{\gamma }^2}\,\xi ),
    \gamma \,\cos (\gamma \,\xi )\,\cos ({{\gamma }^2}\,\xi ) +
   \sin (\gamma \,\xi )\,\sin ({{\gamma }^2}\,\xi ),
  0 \right\}  \label{lam2} \\
 & & \nonumber \\
\lambda_{(3)}^i & = &  \left\{ 0, 0,0,1 \right\} \label{lam3}
\end{eqnarray}
Equations (\ref{lam0})--(\ref{lam3}) are valid for all times $s$
because they are exact solutions of Eqs.\
(\ref{FDiffEq0})--(\ref{FDiffEq3}) and satisfy the orthogonality
relations in Eq.\ (\ref{initialTetrad}).

\section{Fermi Coordinates}

Fermi coordinates are defined by the geometric construction shown
in Figure \ref{ObserverWorldLineFigure}. Every event $P^\prime$ in space-time has coordinates
$x^{i \, \prime}$ in the global inertial coordinate system.
According to the observer moving in circular motion along a
time-like world line,  the same event has the Fermi coordinates
$X^{(a)}$, $a=0,1,2,3$.  The first Fermi coordinate, $X^{(0)}=s$,
is just the proper time (in units of length) associated with the
event $P^\prime$. The proper time for $P^\prime$ is defined as the
value of arc length $s$ in space-time at which a space-like
geodesic from point $P$ (which defines $s$) and passes through
event $P^\prime$. The definition of $s$ requires  the tangent
vector of this space-like geodesic, $\mu^i$, be orthogonal to the
observer 4-velocity at $P$,
\begin{equation}
\mu^i \, v_i \vert_P = 0
\label{normal1}
\end{equation}
For the simple case of Minkowski space, the geodesic is simply a
straight line connecting the points $P \, P^\prime$
\begin{equation}
\mu^i = N ( x^{i \, \prime} - x^i(\gamma \, s) )
\label{muDef}
\end{equation}
where $x^i(u)$ and $x^{i \, \prime}$ are the coordinates of $P$ and
$P^\prime$, respectively, in the global inertial frame, and $N$ is
a normalization constant that makes $\mu^i$ a unit vector. The
orthogonality condition in Eq.\ (\ref{normal1}) is
\begin{equation}
g_{i j} \, \mu^i (s) \, \lambda^j_{(0)}(s) = 0
\label{normalCondition}
\end{equation}
and gives an implicit equation for $s$ for a given $P^\prime$.
This orthogonality condition gives the first Fermi coordinate of
the point $P^\prime$
\begin{equation}
X^{(0)}=s
\label{orthogCondition}
\end{equation}
and Eq.\ (\ref{normalCondition}) gives $s$ as an implicit
equation:
\begin{equation}
\gamma \, s =  x^{0 \, \prime}-x_0^0 + \nu  \left[  (x^{1 \, \prime} - x_0^1) \,\sin (\frac{ \gamma \,\omega \, s }{c})
   - (x^{2 \, \prime} - x_0^2) \, \cos (\frac{\gamma \,\omega \, s}{c}) \right]
\label{sDef}
\end{equation}
In the limit of small speeds of the observer, $\nu \rightarrow 0$, Eq.\
(\ref{sDef}) gives $s=x^{0 \, \prime}-x_0^0$. For  $\nu <<1$, Eq.\
(\ref{sDef}) can be solved for $s$ by iteration, resulting in  a
power series in $\nu$
\begin{eqnarray}
X^{(0)} & = \, s \,= & \Delta x^0 + \left[ \Delta x^1 \, \sin( \frac{\Delta x^0 \,\omega }{c})
  -\Delta x^2 \, \cos ( \frac{\Delta x^0 \,\omega }{c} ) \right]  \,\nu
    - \frac{1}{2}\Delta x^0 \,\nu^2  \nonumber \\
    &  &
     -\frac{1}{2a} \left[ -a + 2\,\Delta x^1 \,
         \cos (\frac{\Delta x^0 \,\omega }{c}) +
        2\, \Delta x^2 \,
         \sin ( \frac{\Delta x^0 \,\omega }{c}) \right] \,  \left[ \Delta x^2 \,
         \cos ( \frac{\Delta x^0 \,\omega }{c}) -
        \Delta x^1 \, \sin ( \frac{\Delta x^0 \,\omega }{c}) \right] \,
      {\nu }^3 + O(\nu )^4
\label{sExplicit}
\end{eqnarray}
where I use the notation
\begin{equation}
\Delta x^i=x^{i \, \prime} - x^i_0
\end{equation}
For future reference, I label the coordinates of $P$, $P^\prime$,
and $P_0$ in the global inertial frame by
$P^\prime  =( x^{0 \, \prime}, x^{1 \, \prime}, x^{2 \,
\prime}, x^{3 \, \prime} )$,
$P =( x^{0}, x^{1}, x^{2}, x^{3} )$, and
$P_0 =( x^{0}_0, x^{1}_0, x^{2}_0, x^{3}_0 ) $.

The contravariant spatial Fermi coordinates, $X^{(\alpha)}$,
$\alpha=1,2,3$, are defined to be\cite{Synge1960}
\begin{equation}
X^{(\alpha)}= \sigma \mu^i  \lambda^{(\alpha)}_i = g_{i j} \,
\sigma(s)  \, \mu^i(s) \, \eta^{(\alpha \beta)} \,
\lambda^j_{(\beta)}(s) \label{FCcoordDef}
\end{equation}
where I used the Minkowski definition of space-time distance, $\sigma$, along the
space-like geodesic between  $P$ and $P^\prime$
\begin{equation}
\sigma^2 = g_{ij}( x^{i \, \prime} - x^i) ( x^{j \, \prime} - x^j)
\end{equation}
where $g_{ij}$ is the Minkowski metric and  $\eta^{(i j)}=\eta_{(i
j)}=\eta_{ij}$ is an invariant matrix, which is also equal to the
Minkowski metric, given in the Appendix. The parameter $s$ in Eq.\
(\ref{FCcoordDef}) is a function of $P^\prime$ coordinates
$x^{i\prime}$, so $s$ must be eliminated using  Eq.\
(\ref{sExplicit}), so that $X^{(\alpha)}$ are explicit functions
of the global inertial coordinates $x^{i\prime}$.  I have
calculated this transformation correct to fourth order in $\nu$.
However, since the expressions are complicated, I write them here
correct only to third order in $\nu$:
\begin{eqnarray}
X^{(1)} & = &
 \Delta x^1 - a\, \cos ( \frac{\Delta x^0 \,\omega }{c})
  + \frac{1}{4} \left[ \Delta x^1 - \Delta x^1 \, \cos ( \frac{2\,\Delta x^0 \,\omega
  }{c})
  -\Delta x^2 \,  \sin (\frac{2\, \Delta x^0 \,\omega }{c}) \right] \,
  \nu^2   \nonumber \\
  &  &  + \frac{1}{2a} \Delta x^0 \, \left[ - \Delta x^2 +
        a \,\sin (\frac{\Delta x^0\,\omega }{c}) \right] \,
        \nu^3  + O(\nu )^4  \label{X1} \\
X^{(2)} & = & \Delta x^2 - a\,\sin ( \frac{\Delta x^0 \,\omega
}{c})    + \frac{1}{4}\left[ \Delta x^2 + \Delta x^2 \,
         \cos (\frac{2\,\Delta x^0\,\omega }{c}) -
         \Delta x^1 \, \sin (\frac{2\,\Delta x^0 \,\omega }{c}) \right] \, \nu^2   \nonumber \\
 &  & + \frac{1}{2 a} \, \Delta x^0 \,
      \left[ \Delta x^1 -  a\,\cos ( \frac{\Delta x^0\,\omega }{c}) \right] \,\nu^3
 + O(\nu )^4
 \label{X2} \\
X^{(3)} &  = &  \Delta  x^3 \label{X3}
\end{eqnarray}
The inverse transformation can be obtained from Eqs.\
(\ref{X1})--(\ref{X3}) by iteration resulting in:
\begin{eqnarray}
x^{0 \, \prime} - x^0_0  & = &
X^{(0)} +  \left[  X^{(2)} \,\cos ( \frac{ \omega X^{(0)}}{c}) -
 X^{(1)} \,\sin (\frac{\omega  X^{(0)} }{c}) \right] \nu
  + \frac{1}{2}\, X^{(0)} \nu^2 + O(\nu )^3 \\
x^{1 \, \prime} - x^1_0  & = & X^{(1)} + a\,\cos (\frac{\omega
X^{(0)}}{c}) + \frac{1}{4} \left[ X^{(1)} - X^{(1)} \,\cos (
\frac{2\, \omega X^{0)}}{c}) -
 X^{(2)} \,\sin (\frac{2\,\omega  X^{(0)} }{c}) \right] \, \nu^2  + O(\nu )^3 \\
x^{2 \, \prime} - x^2_0  & = & X^{(2)} + a\,\sin (\frac{\omega
X^{(0)}}{c})  +\frac{1}{4} \left[ X^{(2)} + X^{(2)} \,\cos (
\frac{2\, \omega X^{(0)}}{c}) -
 X^{(1)} \,\sin (\frac{2\,\omega  X^{(0)} }{c}) \right] \, \nu^2
 + O(\nu )^3 \\
x^{3 \, \prime} - x^3_0  & = & X^{(3)}
\label{inverseTransformation}
\end{eqnarray}

\section{Metric in Fermi Coordinates}

The line element in the Fermi coordinate system of the observer is
written as
\begin{equation}
ds^2=-G_{(ij)} \, dX^{(i)} \, dX^{(j)}
\label{FermiCoordLineElement}
\end{equation}
where $G_{(ij)}$ are the metric tensor components when the
$X^{(i)}$ are used as coordinates. A direct calculation of the
metric tensor components from the tensor transformation rule
\begin{equation}
G_{(ij)}= g_{kl}\, \frac{\partial x^k}{\partial X^{(i)}} \,
 \frac{\partial x^l}{\partial X^{(j)}}
\end{equation}
using the transformation Eq.\ (\ref{X1})--(\ref{X3}) and the
Minkowski metric for $g_{ij}$, gives
\begin{equation}
G_{(00)} = -(1+\zeta)^2 \label{FCmetric}
\end{equation}
and $G_{(\alpha \beta)} = \delta_{\alpha \beta}$ where $\zeta$ is
given by
\begin{equation}
\zeta =
 \left[ 1- \frac{\Delta x^1}{a} \cos ( \frac{\omega \Delta x^0}{c})  -
\frac{\Delta x^2}{a} \sin ( \frac{\omega \Delta x^0}{c})
\right] \, \nu^2 + O(\nu^4)  \\
\label{tildezeta}
\end{equation}
Using the transformation Eqs.\ (\ref{X1})--(\ref{X3}) to express
$\zeta$ as a function of the Fermi coordinates, I obtain:
\begin{equation}
\zeta =  -\frac{1}{a} \left[ X^{(1)} \cos (\frac{\omega}{c}
X^{(0)}) + X^{(2)} \sin (\frac{\omega}{c} X^{(0)})     \right] \,
\nu^2 + O(\nu^4) \label{zetaDef}
\end{equation}

The result in Eq.\ (\ref{FCmetric} ) agrees with Synge's general
theory\cite{Synge1960} of Fermi coordinates specialized to flat
space.   Synge shows that
\begin{equation}
\zeta= X_{(\beta)} \, w^{(\beta)}=
\eta_{(\alpha \beta)} \, X^{(\alpha)}\, w^{(\beta)} =
\eta_{(\alpha \beta)} \, X^{(\alpha)}\, w^{i} \, \lambda^{(\beta)}_i
\label{zetaDef1}
\end{equation}
To third order in $\nu$, Eq.\ (\ref{zetaDef}) agrees with the
general theory given by Synge\cite{Synge1960}.
The quantities $X_{(\beta)}=\eta_{(\alpha \beta)} \, X^{(\alpha)}$
are the covariant Fermi coordinates and $w^{(\beta)} = w^{i} \,
\lambda^{(\beta)}_i$ are the components of the observer's
4-acceleration in the Fermi coordinate system. The
4-acceleration, $w^{i}$, $i=0,1,2,3$, is related to the ordinary
acceleration, $a^\beta$, $\beta=1,2,3$, by
\begin{equation}
w^{i} = \frac{\gamma^2}{c^2} \, \frac{d^2 x^i}{dt^2}=
\frac{\gamma^2}{c^2} \, a^i
\end{equation}
Using this relation, $\zeta$ can be written in terms of the global
inertial coordinates of $P$ and $P^\prime$ as
\begin{equation}
\zeta = \frac{\gamma^2}{c^2} \, \delta_{\alpha \beta}\,
\left( x^{\alpha \, \prime} - x^\alpha (s) \right) \, a^\beta
\label{zetaDef2}
\end{equation}

For the circular motion treated here, from Eq.\ (\ref{zetaDef}), the
three spatial components of the observer's acceleration in
Fermi coordinates are given by
\begin{equation}
w^{(\beta)} =  -\frac{\nu^2}{a} \left( \cos (\frac{\omega}{c}
X^{(0)}), \sin (\frac{\omega}{c} X^{(0)}), 0  \right)
\label{aDef}
\end{equation}
where we have dropped fourth-order terms in $\nu$. To third order
in $\nu$, the observer's acceleration is just the classical
centripetal acceleration for an observer moving in a circle.

The quantity $c^2 \, \zeta$ is a potential (analogous to a
gravitational potential) that determines the rate of proper time
with respect to coordinate time in Fermi coordinates. Proper time
is kept by an ideal clock. Coordinate time depends on the
definition of the reference frame and coordinates used within that
frame. The quantity $\zeta$, given in Eq.\ (\ref{zetaDef2}),
depends on acceleration $a^\beta$, and the difference in
coordinates, $x^{\alpha \, \prime} - x^{\alpha}(s)$, from the
spatial origin of the Fermi coordinates. Away
from the spatial origin of Fermi coordinates the acceleration
produces an effective gravitational potential field, $c^2 \,
\zeta$, which leads to a change in the relation of coordinate time
to proper time, through the metric components $G_{(ij)}$.

\section{Apparent Behavior of Clocks in Fermi Coordinates}

The significance of proper time is that it is the quantity that
ideal clocks keep, that it may be related to time on real clocks,
and that it is closely related to measurable quantities. The
significance of coordinate time is that it enters into the theory.
In order to analyze experiments with clocks, we must relate what
is measured, i.\ e.\ proper time intervals between space-time
events, to coordinate positions and times of these events.

Consider an ideal clock whose world line is given in Fermi
coordinates by $X^{(i)}=X^{(i)}(X^{(0)})$.  At coordinate times
$X^{(0)}_A$, and $X^{(0)}_B$, the clock is at spatial positions
$X^{(\alpha)}_A$, and $X^{(\alpha)}_B$, respectively. The elapsed
proper time between these two events, $\Delta \tau=\Delta s/c$, is
given by the integral along the world line of the clock
\begin{equation}
\Delta s = \int_A^B \left( -G_{(ij)}  \frac{dX^{(i)}}{dX^{(0)}} \,
\frac{dX^{(j)}}{dX^{(0)}} \right)^{1/2} \, dX^{(0)}
\label{worldLineIntegral}
\end{equation}

\subsection{Stationary Clock}

For the case of a stationary clock, located at constant Fermi
coordinate position $X^{(\alpha)}$,  the relation between proper
time $\tau$ and coordinate time, $X^{(0)}$, is given by taking
$dX^{(\alpha)}=0$ in Eq.\ (\ref{FermiCoordLineElement}).  The {\it
rate} of proper time with respect to coordinate time is then given
by
\begin{equation}
\frac{ds}{d X^{(0)}} =  1 + \zeta( X^{(0)}, X^{(\alpha)})
\label{StationaryClockRate}
\end{equation}

Equation (\ref{StationaryClockRate}) gives the well-known result
that the  rate of proper time depends on the location of the
clock~\cite{LLClassicalFields}. For the simple case of circular
motion of the observer with $\zeta$ given by Eq.\ (\ref{zetaDef}),
substitution in Eq.\ (\ref{StationaryClockRate}) and integration from $X^{(0)}_A$
to $X^{(0)}_B$ gives
\begin{eqnarray}
\Delta s & = &  X^{(0)}_B - X^{(0)}_A - \frac{a \omega}{c}
\left\{  X^{(1)} \, \left[  \sin \left(
\frac{\omega}{c}X^{(0)}_B \right) - \sin \left(
\frac{\omega}{c}X^{(0)}_A \right) \right]  -X^{(2)}\left[  \cos
\left(   \frac{\omega}{c}X^{(0)}_B \right) - \cos \left(
\frac{\omega}{c}X^{(0)}_A \right) \right] \right\}
\label{StationaryClock}
\end{eqnarray}
where I have dropped terms of $O(\nu^4)$. Equation
(\ref{StationaryClock}) shows that there is a complicated relation
between elapsed proper time on a clock located at Fermi coordinates
$X^{(1)},X^{(2)}$, and Fermi coordinate time, $X^{(0)}$.  For
example, a clock on the $X^{(1)}$ axis (with $X^{(2)}=0$) has a
periodic time with respect to Fermi coordinate time, given by
$\Delta s = X^{(0)}_B - (a \omega/c) \, X^{(1)} \, \sin(\omega
X^{(0)}_B/c)$. The initial conditions in Eq.\
(\ref{StationaryClock}) determine complicated phase relations
between proper time and coordinate time.

\subsection{Clock in Circular Motion}

Now consider an ideal clock that is a distance $b$ from the origin
of Fermi coordinates and the clock executes circular motion in
Fermi coordinates in the $X^{(1)}-X^{(2)}$ plane. I take the world line
for this motion to be given by
\begin{eqnarray}
X^{(1)} & = &   b \cos \left( \frac{\Omega}{c} X^{(0)} \right)\label{circularClockWorldLine1} \\
X^{(2)} & = &   b \sin \left( \frac{\Omega}{c} X^{(0)} \right) \label{circularClockWorldLine2}\\
X^{(3)} & = &  0
\label{circularClockWorldLine3}
\end{eqnarray}
The clock moves in the $X^{(1)}$-$X^{2)}$ plane, which coincides
with the plane defined by the circular motion of the origin of
Fermi coordinates. See Figure \ref{ObserverWorldLineFigure}.  The
situation considered here is a simplification of the problem of a
GPS satellite  (with a clock) orbiting the Earth, which is itself
orbiting the sun.  (In the case of GPS, the satellites are in
orbits that are inclined at 55$^o$ with respect to the Earth's
equator.) However, as mentioned previously, in the case of the
Earth and GPS satellite, both travel on geodesics. Since we are
neglecting the effects of gravity in the problem considered here,
the Fermi observer and orbiting clock are both Fermi-Walker
transported, e.\ g., with the aid of rockets.

Substitution of the satellite clock's world line into Eq.\
(\ref{worldLineIntegral}), leads to the following relation between
elapsed proper time observed on the clock and Fermi coordinate time
\begin{eqnarray}
\Delta s & = & \left[1-\frac{1}{2}\left(\frac{b \,
\Omega}{c}\right)^2 \right] ( X^{(0)}_B - X^{(0)}_A ) -\frac{a \,
b}{c}\frac{\omega^2}{\Omega-\omega}\left[ \sin
\left(\frac{\Omega-\omega}{c}X^{(0)}_B \right) -
\sin\left(\frac{\Omega-\omega}{c}X^{(0)}_A \right) \right]
\label{orbitingClock1}
\end{eqnarray}
The {\it rate} of proper time with respect to coordinate time is
\begin{equation}
\frac{d s}{d  X^{(0)}}  =  -\frac{1}{2}\left(\frac{b \, \Omega}{c}\right)^2
 -\left(\frac{\omega}{c}\right)^2 a \, b  \cos
\left(\frac{\Omega-\omega}{c}X^{(0)} \right)
\label{orbitingClock1Rate}
\end{equation}

The proper time has a constant rate offset from coordinate time,
represented by the first term in Eq.\ (\ref{orbitingClock1}). In addition, the
proper time has a periodic component with respect to coordinate time, given by
the second term in \mbox{Eq.\ (\ref{orbitingClock1})}.

To demonstrate the magnitude of these effects, I use parameters
that correspond to Earth-Sun distance and GPS satellites. I
compute the magnitudes of the following terms with the values
shown in Table \ref{ConstantValues}:
\begin{equation}
\frac{1}{2} \,\left(\frac{b \, \Omega}{c}\right)^2 = 8.348\times
10^{-11} \label{TimeDilationRateOffset}
\end{equation}
\begin{equation}
\frac{1}{2} \,\left(\frac{b \, \Omega}{c}\right)^2 \,
\times\,\frac{2 \pi}{\Omega} =
 3.597\times 10^{-6}  \; {\rm sec}
 \label{TimeDilationOffset}
\end{equation}
Equation (\ref{TimeDilationRateOffset}) gives the constant rate
offset of the moving clock with respect to Fermi coordinate time,
due to time dilation. In GPS, this term leads to a net slowing
down of GPS clocks by approximately 7 $\mu$s per
day\cite{AshbySpilkerBook}, which results from multiplying the
value given in Eq.\ (\ref{TimeDilationOffset}) for a single
revolution by a factor of 2 (since the GPS satellites make
approximately two revolutions per Earth day.)

The second term in Eq.\ (\ref{orbitingClock1}) is a
periodicity of the proper time with respect to Fermi coordinate time
with amplitude
\begin{equation}
\frac{a b}{c} \frac{\omega^2}{\Omega-\omega} =
12.05 \times 10^{-9} \; {\rm sec}
\end{equation}
which corresponds to an amplitude in the rate
\begin{equation}
a \, b \left(\frac{\omega}{c}\right)^2 = 1.755\times 10^{-12}  \;
{\rm sec}
\end{equation}

The proper time given by Eq.\ (\ref{orbitingClock1}) is periodic
with respect to Fermi coordinate time with the difference
frequency, $\Omega - \omega$, which is the difference frequency
between satellite and observer rotations. This is a kinematic
effect due to the type of coordinate system used (Fermi
coordinates) and the fact that the observer (Fermi coordinate
frame) is moving along an arc of a circle, and therefore
experiences an acceleration. Note that the periodic term vanishes
when the observer angular velocity $\omega=0$. If the observer
were moving a straight line and hence have zero acceleration,
$w^i=0$, the periodic effect would also be absent, as can be seen
from Eq.\ (\ref{zetaDef1}).

\subsection{Sagnac-Like Effect for Portable Clocks}

Sagnac\cite{Sagnac,SagnacEffectReview,RobertsonNoonan1968}
demonstrated that there exists a phase shift between two counter
propagating light beams on a rotating platform and that this
phase shift depends on the angular frequency of rotation of the
platform. In the Fermi coordinates of the observer moving in a
circle, there exists a Sagnac-like effect for two portable clocks
that are orbiting in opposite directions.

Consider a clock that moves eastward along a circle of radius $b$
in the equatorial plane of the Fermi coordinates with world line
given by Eq.\
(\ref{circularClockWorldLine1})--(\ref{circularClockWorldLine3}).
See Figure \ref{ObserverWorldLineFigure}. The speed of the clock in Fermi coordinates is $b
\Omega/c$. Assuming that the clock begins its journey at point $A$
at time $X^{(0)}_A=0$ and ends it at $X^{(0)}_B=2 \pi c/\Omega$,
which corresponds to one revolution, the proper time that elapses
on the clock is given by Eq.\ (\ref{orbitingClock1}) as
\begin{equation}
\Delta s_+  = \left[1-\frac{1}{2}\left(\frac{b \, \Omega}{c}\right)^2
\right] \frac{2 \pi}{\Omega}c
+ \frac{a \, b}{c}\frac{\omega^2}{\Omega-\omega} \sin
\left(2 \pi \frac{\omega}{\Omega} \right) \label{rightClock}
\end{equation}
The first term in Eq.\ (\ref{rightClock}) is the time dilation due
to the speed $b \Omega /c$ of the clock.  The second term depends
on the path traversed.

Next, consider an identical clock that starts at point $A$ at time
$X^{(0)}_A=0$ but moves westward. The world line of this clock is
then given by Eq.\
(\ref{circularClockWorldLine1})--(\ref{circularClockWorldLine3})
but with $\Omega=-|\Omega|$. The proper time on this clock is
given by
\begin{equation}
\Delta s_-  = \left[1-\frac{1}{2}\left(\frac{b \, \Omega}{c}\right)^2
\right] \frac{2 \pi}{|\Omega|}c
- \frac{a \, b}{c}\frac{\omega^2}{|\Omega|+\omega} \sin
\left(2 \pi \frac{\omega}{|\Omega|} \right) \label{leftClock}
\end{equation}
When the east-moving and west-moving clocks have each made one
revolution and they have returned to their starting positions at point $A$,
the difference in their times is given by $\Delta  \tau_+ -  \Delta \tau_- =
(\Delta s_+ - \Delta s_- )/c $. For the case applicable to GPS,
$b << a$ and $\omega << |\Omega|$, the difference on the two
clocks is given by
\begin{eqnarray}
 \tau_+ - \tau_- & = &  2 \frac{b}{a} \left(\frac{a \omega}{c}
 \right)^2  \frac{|\Omega|}{\Omega^2 - \omega^2} \sin\left( 2\pi
 \frac{\omega}{| \Omega |} \right) \\
  &  = & 4\pi \frac{b}{a} \left(\frac{a \omega}{c}
 \right)^2 \frac{\omega}{\Omega^2 } + O(\frac{\omega^2}{\Omega^2})
\label{eastWestTimeDifference}
\end{eqnarray}
where in the last line I have dropped terms of
$O(\omega^2/\Omega^2)$. This Sagnac-type effect is due to the
non-inertial nature of the Fermi coordinate system. The origin of
this effect can  be understood from the point of view of the
global inertial coordinate system. If the Fermi coordinate origin
moved along a straight line in Minkowski space (rather than a
circle) then there would be no acceleration and $\zeta=0$. In this
case, the east-moving and west-moving clocks would have zero time
difference after one revolution. However, since the origin of
Fermi coordinates moves along a circle, the acceleration picks out
a direction in the Fermi coordinate space-time and the symmetry
between the two clock world lines is broken. Consequently, the
elapsed proper time is different on the two clocks when they are
compared after one revolution. Note that the time difference on
the two clocks increases with increasing angular velocity of
rotation $\omega$, analogous to the optical Sagnac
effect\cite{Sagnac,SagnacEffectReview,RobertsonNoonan1968}. This
increase is due to the fact that the breaking in symmetry is due
to the magnitude of the acceleration, $w^{(\beta)}$, see Eq.\
(\ref{zetaDef1}). Using the values of the parameters appropriate
to GPS satellites, given in Table 1, I find the magnitude of the
time differernce on the two clocks after one revolution to be
\begin{equation}
 \tau_+ - \tau_-  = 2.066 \times 10^{-10} \; \; \; {\rm sec}
\label{eastWestTimeDifferenceNumber}
\end{equation}
This is a small effect that vanishes when the origin of Fermi
coordinates moves along a straight line or when $\omega=0$.

\section{Coordinate Speed of Light}

The three dimensional space of Minkowski space-time is homogeneous
and isotropic, and consequently, the speed of light is a constant.
However, for the case of Fermi coordinates of an observer moving
in a circle, the symmetry of the 3-d space is broken by an
acceleration. Consequently, the space is not homogeneous and
isotropic and the coordinate speed of light depends on position in
the 3-d space.

\begin{figure}[htbp]
\centerline{\epsfsize=6.0cm \epsfbox{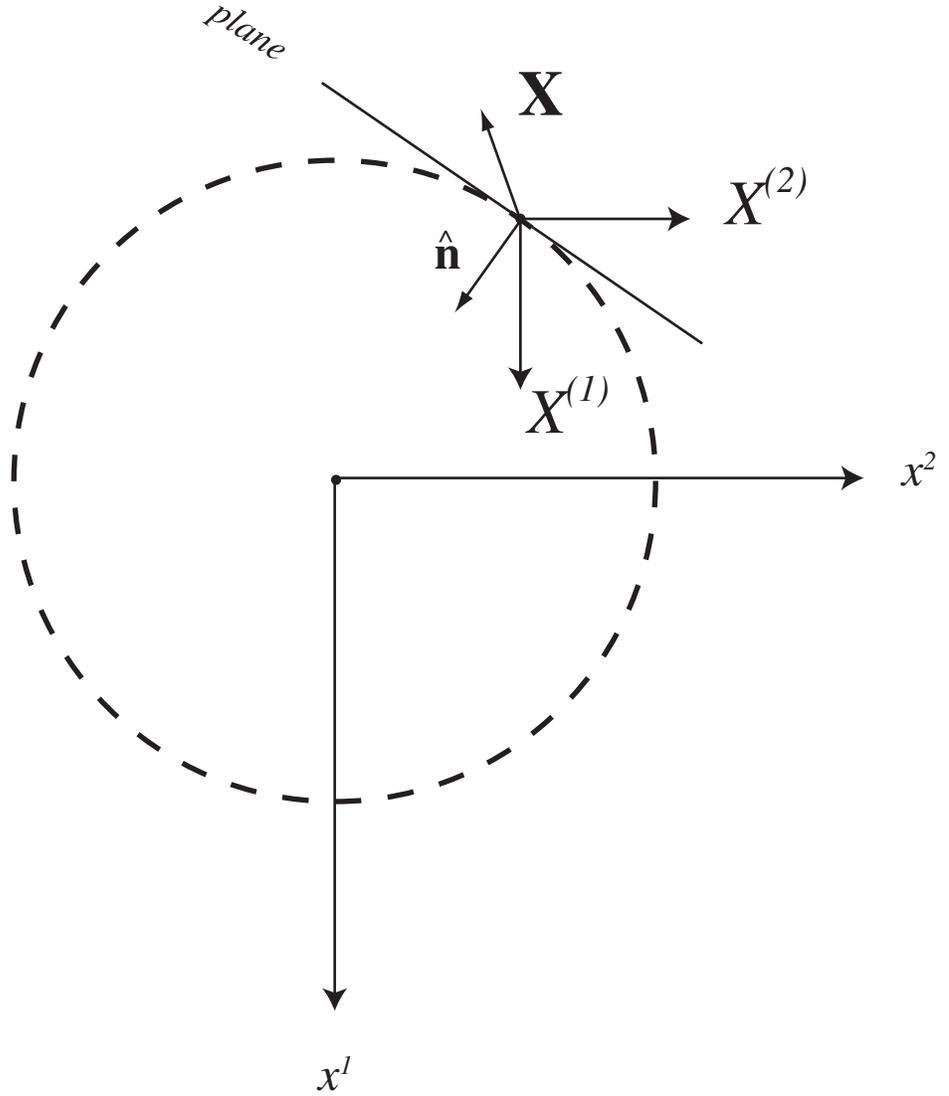}}
\caption{The global inertial coordinate axes $x^i$ and
Fermi coordinates $X^{(a)}$ are shown in relation to the coordinate
position ${\bf X}$, unit vector ${\bf \hat{n}}$, and the plane
${\bf X}\cdot {\bf \hat{n}}=0$, at a given coordinate time
$X^{(0)}$. } \protect \label{SpeedOfLightFigure}
\end{figure}

For a general space-time metric $g_{ij}$, the coordinate speed of
light $v_c$ in the direction of a unit vector $e^\alpha$ is given
by\cite{Moller1972}
\begin{equation}
v_c \, = \, \frac{\sqrt{-g_{00}}}{1+\gamma_\alpha e^\alpha} \, c
\label{lightSpeed}
\end{equation}
where
\begin{equation}
\gamma_\alpha =\frac{g_{0\alpha}}{\sqrt{-g_{00}}}
 \label{gammaVecDef}
\end{equation}
and the unit vector satisfies the relation $\gamma_{\alpha \beta}
\, e^\alpha \, e^\beta =1$, where the 3-d spatial metric is given
by\cite{LLClassicalFields}
\begin{equation}
\gamma_{\alpha \beta}=g_{\alpha \beta} - \frac{g_{0\alpha} \,
g_{0\beta}}{g_{00}}
\end{equation}
Equation (\ref{lightSpeed}) shows that for a general space-time
metric the speed of light depends on direction $e^\alpha$ and on
position and time, through the metric components. A non-zero
off-diagonal metric component $g_{0\alpha}$ leads to an anisotropy
(directional dependence) of the coordinate speed of light. The
metric for Fermi coordinates given in Eq.\ (\ref{FCmetric}) has
vanishing off-diagonal terms,$\gamma_\alpha=0$, and consequently
the coordinate speed of light is isotropic. However, the metric
component $G_{(00)}$ has a non-trivial position and time
dependence, and Eq.\ (\ref{lightSpeed}) gives the coordinate speed
of light:
\begin{equation}
v_c = c \, (1 + \zeta)
 \label{FClightSpeed}
\end{equation}
where $\zeta$ is a function of Fermi coordinate position and time,
given by Eq.\ (\ref{zetaDef}). Therefore, I can write the
coordinate dependence of the speed of light as
\begin{equation}
\frac{\Delta c}{c} =  -\frac{\nu^2}{a} \left[ X^{(1)} \cos
(\frac{\omega}{c} X^{(0)}) + X^{(2)} \sin (\frac{\omega}{c}
X^{(0)})     \right] + O(\nu^4)
 \label{FClightSpeedVariation}
\end{equation}
This variable speed of light is a kinematic effect due to the
non-inertial nature of the Fermi coordinate system. Note that at
any given position ${\bf X}=\{X^{(1)},X^{(2)},X^{(3)} \}$ the
speed of light depends on time in a periodic way, with the orbital
period of the reference frame, $\omega$. This expression can be
rewritten as
\begin{equation}
\frac{\Delta c}{c} =  a\, \left(\frac{\omega}{c}\right)^2 {\bf
X}\cdot {\bf \hat{n}} + O(\nu^4) \label{FClightSpeedVariation2}
\end{equation}
where the unit vector ${\bf
\hat{n}}=\left(-\cos\left(\frac{\omega}{c}X^{(0)}\right),
-\sin\left(\frac{\omega}{c}X^{(0)}\right),0 \right)$ is in the
observer's orbital plane and points in the direction of
acceleration,  toward the center of the circle. See Figure
\ref{SpeedOfLightFigure}. As the Fermi reference frame moves
around the circle, the vector  ${\bf \hat{n}}$ always points
toward the circle's center in Fermi coordinates. At any given time
$X^{(0)}$, Eq.\ (\ref{FClightSpeedVariation2}) shows that to
second order in $\nu$, in the plane given by
\begin{equation}
{\bf X}\cdot
{\bf \hat{n}}=0 \label{plane}
\end{equation}
the speed of light does not differ from it's vacuum inertial frame
value $c$. In the 3-d half-space containing  positions ${\bf X}$
closer to the center of the circle, the speed of light is
increased.  For the 3-d half-space containing positions ${\bf X}$
that are further from the center of the circle, the speed of
light is decreased. As time $X^{(0)}$ increases, the dividing
plane, given by Eq.\ (\ref{plane}), rotates at angular frequency
$\omega$ around the circle, with the origin of Fermi coordinates
remaining the point at which the plane is tangent to the circle.

The magnitude of the inhomogeneous variation in the speed of light
in Eq.\ (\ref{FClightSpeedVariation}) is small.  For the
kinematics of GPS, using the numbers given in Table 1, with the
coordinates $X^{(1)}$, $X^{(2)}\sim b$, so the effect has a
magnitude
\begin{equation}
\frac{\nu^2}{a}\, b = 1.755\times 10^{-12}
\end{equation}
and has a complicated position and time dependence given by Eq.\
(\ref{FClightSpeedVariation}).  Recently, GPS data have been used
to obtain an upper bound on the {\it anisotropy} of the speed of
light to be $\Delta c/c < 10^{-9}$, see Ref.~\cite{Wolf97} and
references contained therein.  Given the present accuracy of GPS,
the small inhomogeneity of the speed of light, given by Eq.\
(\ref{FClightSpeedVariation})is probably not measurable.

\section{Summary}

I have considered the problem of an observer moving in a circle
around a central point in flat space (no gravitational fields).
The observer carries with him three gyroscopes and uses them to
define a set of non-rotating axes.  He keeps his axes at a
constant orientation with respect to the gyroscopes' spin axes.
The observer's axes are Fermi transported along his world line and
Fermi coordinates are defined.  The Fermi time coordinate is just
the observer's proper time, according to the construction shown in
Figure \ref{ObserverMotionFigure}.  Fermi coordinates play the
role of ECI frame coordinates in GPS.  I investigate the kinematic
effects seen on ideal clocks from the point of view of these Fermi
coordinates.  Specifically, I consider two cases: stationary
clocks (with respect to Fermi coordinates) and satellite clocks in
circular orbit with respect to Fermi coordinates. The proper time
interval between two events on the stationary clock, given by Eq.\
(\ref{StationaryClock}), is a complicated periodic function of the
clock's Fermi coordinate time and coordinate position in the
plane perpendicular to the rotation axis. I considered
next a clock that orbits the Fermi coordinate origin (similar to a
satellite) in the equatorial plane of rotation and computed the
relation of proper time to coordinate time. I find that the proper
time on the clock has time dilation (constant rate offset) plus a
periodic dependence on Fermi coordinate time at the difference
frequency, $\Omega-\omega$, between satellite clock orbital
frequency, $\Omega$, and Fermi coordinate frame orbital frequency,
$\omega$ .

Next, I looked at the time difference of two counter-orbiting clocks
in the equatorial plane of rotation, one west-moving and the other
east-moving. After one revolution, the difference in proper time
on the clocks shows a small Sagnac-like effect, increasing with
the angular frequency of revolution of the Fermi frame, $\omega$,
given by Eq.\ (\ref{eastWestTimeDifference}) .

Finally, I computed the coordinate speed of light in the Fermi
coordinates.  The speed of light is isotropic (same in all
directions), but is not homogeneous.  Specifically, the speed of
light depends on Fermi coordinate position and time, but not on
direction.  Equation (\ref{FClightSpeedVariation2}) gives the
change in the speed of light from $c$, its vacuum value in an
inertial frame.

I have computed the magnitudes of these effects for numbers
relevant to GPS satellite orbits and orbital frequency. The
effects are all small for GPS, except for the 12 ns periodicity
for an orbiting clock, given in Eq.\ (\ref{orbitingClock1}), that
occurs at the difference frequency, $\Omega-\omega$, between
satellite orbital frequency $\Omega$ and Earth orbital frequency
$\omega$. This effect is reminiscent of the periodic effect in GPS
described in the introduction. However, the effects here are all
kinematic since gravitational fields are neglected. The inclusion
of gravitational fields is expected to significantly modify the
results.  In particular, from the work of Ashby and Bertotti, this
12 ns periodic effect is expected to cancel when gravitational
fields are included~\cite{ParkinsonGPSReview,AsbyBertotti}.  Work
is in progress to extend these results to include the effects of
gravitational fields.

While the effects discussed in this work are small for the present
GPS, these effects are quite significant for future navigation and
time transfer systems that are expected to have sub-nanosecond
time transfer accuracy~\cite{AsbyAllan96,PetitWolf94,WolfPetit95}.

\acknowledgments
The author thanks Tom Van Flandern for useful
discussions.

\appendix
\section{Conventions and Notation}

I use the notation and conventions of Synge\cite{Synge1960},
except that my space-time indices on coordinates, $x^i$, take
values $i=0,1,2,3$. Specifically,  I use the convention that Roman
indices take the values $i=0,1,2,3$, and Greek indices
$\alpha=1,2,3$. I use the summation convention for repeated
indices, when the same index appears in a lower and upper position
then summation is implied over the range of the index.

If $x^i$ and $x^i + dx^i$ are two events along the world line of
an ideal clock, then the square of the proper time interval
between these events  is $d\tau=ds/c$, where $ds$ is given in
terms of the space-time metric as
\begin{equation}\label{metricDef}
ds^2 = -g_{ij}\, dx^i \, dx^j
\end{equation}
and I choose $g_{ij}$ to have the signature $+$2. When $g_{ij}$ is
diagonalized at any given space-time point, the elements can take
the form of the flat space Minkowski metric given by $g_{00}=-1$,
$g_{\alpha \beta}=\delta_{\alpha \beta}$, and $g_{0 \alpha}=0$.

In this work, I take the global inertial frame coordinates to have the
Minkowski metric.  The observer's frame is described by Fermi
coordinates, $X^{(a)}$, $a=0,1,2,3$, with line element
\begin{equation}
ds^2 = -G_{(ij)}\, dX^{(i)} \, dX^{(j)}
\label{FermiMetricLineElement}
\end{equation}

\begin{table}
\caption{Numerical values of constants. \label{ConstantValues}}
\begin{tabular}{ccc}
constant  & definition  & value \\ \tableline
 $a$  & Earth orbital semi-major axis         & 1.496$\times$10$^{11}$ m \\
$\omega$ & Earth orbital angular velocity   & 1.99238$\times$10$^{-7}$ sec$^{-1}$ \\
$b$  & GPS satellite semi-major axis & 2.656177$\times$10$^7$ m \\
$\Omega$  & GPS satellite angular velocity & 1.45842$\times$10$^{-4}$ sec$^{-1}$  \\
$c$  & vacuum speed of light  & 2.997924$\times$10${^8}$ m/s
\end{tabular}
\end{table}

\end{document}